\documentstyle[12pt,epsf]{article}
\makeatletter
\@addtoreset{equation}{section}
\makeatother

\newcommand{\err}[2]{{{$(\makebox[0.8em][r]{#2})$}}}
\begin{document}
\thispagestyle{empty}
\noindent \hspace{1cm} June 1993 \hfill HLRZ\,-\,93-43 \hspace{1cm}\\
\mbox{}                          \hfill BI-TP 93/30    \hspace{1cm}\\
\mbox{}                          \hfill FSU-SCRI-93-76 \hspace{1cm}\\
\mbox{}                          \hfill WUB 93-25 \hspace{1cm}\\
\begin{center}
\vspace*{1.0cm}
{\large \bf The Spatial String Tension in the Deconfined Phase \\
            of the (3+1)-Dimensional SU(2) Gauge Theory} \\
\vspace*{1.0cm}
{\large G.S.\,Bali$^1$, J.\,Fingberg$^2$, \\ [4mm]
        U.M.\,Heller$^2$, F.\,Karsch$^{3,4}$ and K.\,Schilling$^1$}
\vspace*{1.0cm}

{\normalsize
\leftline{$\mbox{}^1$ {Fachbereich Physik, Universit\"at-Gesamthochschule
                       Wuppertal,}}
\centerline{Gausstr.\,20, D-42097 Wuppertal, Germany}
\leftline{$\mbox{}^2$ {SCRI, The Florida State University,
                       Tallahassee, FL 32306-44052, U.S.A.}}
\leftline{$\mbox{}^3$ {HLRZ, c/o KFA J\"{u}lich, D-52425 J\"ulich, Germany}}
\leftline{$\mbox{}^4$ {Fak. f. Physik, Univ. Bielefeld, P.\,O.\,Box 100131,
                       D-33501 Bielefeld, Germany}}}
\vspace*{2cm}
{\large \bf Abstract}
\end{center}

\setlength{\baselineskip}{1.3\baselineskip}
We present results of a detailed investigation of the temperature
dependence of the spatial string tension in $SU(2)$ gauge theory.
We show, for the first time, that the spatial string tension is scaling on the
lattice and thus is non-vanishing in the continuum limit.
It is temperature independent
below $T_c$ and rises rapidly above. For temperatures larger than $2T_c$
we find a scaling behaviour consistent with
 $\sigma_s(T) = (0.136\pm 0.011)g^4(T)T^2$,
where $g(T)$ is the 2-loop running coupling constant with a scale
parameter determined as $\Lambda_T = (0.076\pm 0.013)T_c$.
\vskip1.0truecm
\leftline{PACS numbers: 11.15.Ha; 12.38.G}
\newpage
\setcounter{page}{1}
Non-abelian $SU(N)$ gauge theories in (3+1)-dimensions are known to undergo a
deconfining phase transition at high temperature. The physical string tension,
characterizing the linear rise of the potential between static quark sources
with distance, decreases with increasing temperature and vanishes above $T_c$.
The potential becomes a Debye screened Coulomb potential in the high
temperature phase. While the leading high temperature behaviour as well as the
structure of the heavy quark potential for temperatures well above $T_c$ can
be understood in terms of high temperature perturbation theory, it also is
expected that non-perturbative effects like the generation of a magnetic
mass term, $m_m\sim g^2(T)T$, in the gluon propagator may influence the
spectrum in the high temperature phase. These non-perturbative effects in the
magnetic sector will manifest themselves in correlation functions
for the spatial components of gauge fields.

(3+1)-dimensional renormalizable quantum field
theories at high temperature, through dimensional reduction, can be
reformulated as effective 3-dimensional theories, with the scale of the
dimensionful couplings given in terms of the temperature~\cite{Appelquist}.
In the case of an $SU(N)$ gauge theory the effective theory is a 3-dimensional
gauge theory with adjoint matter (Higgs) fields, emerging from the temporal
component of the gauge fields. Basic properties of the gauge invariant
correlation functions for spatial components of the gauge fields - the
spatial Wilson loops - can be understood in terms of this effective
theory. For instance, as this effective theory is confining, it is natural
to expect that spatial Wilson loops obey an area law behaviour in the high
temperature phase
\begin{equation}
W(R,S) = \langle e^{i\oint_{R\times S} dx_\mu A_\mu} \rangle \sim
e^{-\sigma_s RS}~~~~,
\label{ssloop}
\end{equation}
where $\sigma_s$ has been called the spatial string tension,
although one should stress
that it is not related to properties of a physical potential in the
(3+1)-dimensional theory.
In the case of QCD the effective theory itself is quite complicated even
at high temperatures, as the non-static modes do not decouple from the static
sector~\cite{Landsman}. An analysis of
the temperature dependence of the spatial
string tension thus yields information on the importance of the non-static
sector for long-distance properties of high temperature QCD.

The existence of a non-vanishing spatial string tension,
$\sigma_{s}$, in the high
temperature phase of (3+1)-dimensional $SU(N)$ lattice gauge theory can be
proven rigorously at finite lattice spacing~\cite{Borgs}.
However, despite its basic relevance for a better understanding of the
non-perturbative
structure of non-abelian gauge theories at high temperature, little effort has
been undertaken to arrive at a quantitative description of the properties
of the spatial string tension.
First numerical studies~\cite{Manousakis,Berg} suggested that
$\sigma_s$ stays non-zero but temperature independent in the
high temperature phase
of QCD. Some indications for an increase of $\sigma_s$ with
temperature have been
found recently~\cite{PLacock}. So far no study of the scaling behaviour
of the spatial string tension and its behaviour in the continuum limit
exists.

We present here the results of a detailed, high
statistics analysis of the spatial string tension. The finite
temperature $SU(2)$ gauge theory has been simulated on lattices of size
$N_{\tau} \times 32^3$, with $N_{\tau}$ ranging from 2 to 32. The
simulations have been performed at two values of the gauge coupling,
$\beta = 2.5115$ and $\beta =2.74$, which correspond to the critical
couplings for the deconfinement transition on lattices with temporal
extent $N_{\tau}=8$ and $N_{\tau} =16$, respectively~\cite{Fingberg}.
The lattice spacing thus changes by a factor $2.00\pm 0.04$,
where the error is caused by the uncertainty in both of the
critical couplings.
We confirm this factor through a calculation of the
string tension at low temperatures in the confining phase.
On a lattice of size $16 \times 32^3$ ($32^4$) at $\beta=2.5115$ $(2.74)$
we obtain
\begin{eqnarray}
\sqrt{\sigma}a = \cases{ 0.1836\pm 0.0013 &, $\beta =2.5115$ \cr
0.0911\pm 0.0008 &, $\beta =2.74$}~~~,
\label{zerost}
\end{eqnarray}
which corresponds to a change in lattice spacing
$a_{\beta=2.5115}/a_{\beta=2.74} = 2.016\pm 0.023$, and is consistent
with the factor two obtained from the calculation of the
above critical couplings for the deconfinement transition.

We determine the spatial string tension from temperature dependent
{\it pseudo-potentials} constructed from Wilson loops of size $R \times S$,
where both sides of the loop point into spatial directions,
\begin{equation}
V_T (R) = \lim_{S\rightarrow \infty} \ln{W(R,S) \over W(R,S+1)}~~~.
\label{potent}
\end{equation}
In the actual calculation we also construct off-axis loops in spatial
directions and use standard smearing techniques~\cite{ApeWUB} to improve
the convergence of approximants with increasing $S$.

At fixed gauge coupling the temperature can be varied by varying the
temporal extent, $N_{\tau}$, of the lattice. For $\beta=2.74$ we have
studied the {\it pseudo-potentials} on lattices of size $N_\tau =16, 12, 8,
6, 4$ and $2$, which corresponds to temperatures $T/T_c =1, 1.33, 2, 2.67,
4$ and $8$ in addition to the physical potential at ``zero" temperature
on a $32^4$ lattice.
In order to check the scaling behaviour of the spatial string
tension in the continuum limit we have performed additional calculations at
$\beta=2.5115$ and $N_\tau =8, 6$ and $4$, ie. $T/T_c =1, 1.33$ and
$2$ as well as at ``zero" temperature, approximated by a
$16\times 32^3$ lattice.
Note that this procedure induces only an overall error into the
temperature $T/T_c$ from the uncertainty in the scale $T_c$,
stemming from the error in $\beta_c$~\cite{Fingberg}.
Varying $\beta$ at fixed $N_\tau$ to change
the temperature, as has been customary so far, would introduce
additional errors because the relation between the lattice spacing
$a$ and the coupling $\beta$ is not known sufficiently well.

\begin{figure}[htb]
\begin{center}
\leavevmode
\epsfxsize=280pt
\epsfbox[20 30 620 600]{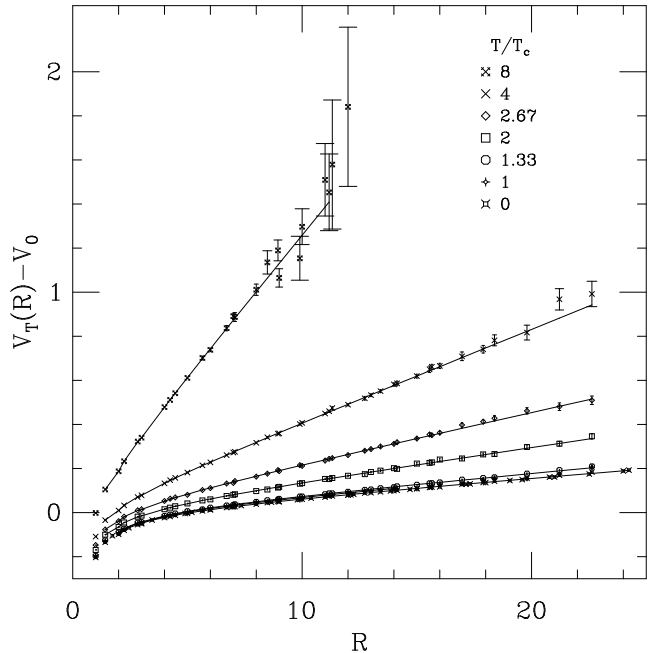}
\end{center}
\caption{\it The pseudo-potentials $V_T(R)$
minus the (constant) self-energy contributions $V_0$ (Eq.~(4))
on lattices of size $N_\tau \times 32^3$ for $\beta=2.74$
as a function of the spatial separation $R$ measured in lattice units.}
\label{fig:eff_pot_all}
\end{figure}

The pseudo-potentials defined through Eq.~(\ref{potent})
are shown in Fig.~\ref{fig:eff_pot_all} for $\beta=2.74$. Obviously
the effective potentials do not show any significant
temperature dependence up to $T_c$. However, as can also be seen from
the figure the slope of the potential rises rapidly above
$T_c$. In order to quantify the temperature dependence of the linearly
rising potentials we follow Refs.~\cite{Michael,Bali} and
fit the potentials with the ansatz
\begin{equation}
V_T(R) = V_0 + \kappa R - {e \over {R}} -
             f \left(G_L(R) - {1 \over {R}} \right)~~~,
\label{fit}
\end{equation}
where $G_L$ denotes the lattice Coulomb potential.
This last term takes account of the lattice artefacts present at small
distances. We have tried
various other fits, including fits where the Coulomb part has been
replaced by a logarithmic term, which would be expected in the high
temperature limit. Details on these fits as well as a
discussion of the short distance part of the pseudo-potentials will be
presented elsewhere. In general we found, that the
fit parameter $\kappa \equiv \sigma_s a^2$,
only weakly depends on the
actual parametrization of the short distance part of $V_T (R)$.
Our results are summarized in table~1.
We determine the spatial string tension in units of the critical temperature,
\begin{equation}
{\sqrt{\sigma_s(T)} \over T_c} = \sqrt{\kappa(T)} N_{\tau,c} ~~~,
\end{equation}
where $N_{\tau,c} = 8$ (16) for $\beta = 2.5115$ (2.74).
These numbers are given in the last column of table~1.

\begin{table}
\begin{center}
\begin{tabular}{|c|c|c|cccc|c|}
\hline
 $T/T_c$ & $\beta$ & meas. & $V_0$ & $\kappa$ & $e$ & $f$ &
 $\sqrt{\sigma_s}/T_c$ \\
\hline
0   & 2.7400 &  835 & .482\err{ 1}{ 3} & .0083\err{ 1}{ 1} & .220\err{ 1}{12}
&
             .13\err{ 1}{ 8} &1.46\err{ 1}{ 1} \\
1   &        &  918 & .475\err{ 2}{ 6} & .0089\err{ 2}{ 6} & .210\err{ 5}{19}
&
             .13\err{ 3}{12} &1.51\err{ 2}{ 5} \\
1.33&        &  720 & .474\err{ 1}{ 3} & .0094\err{ 1}{ 2} & .207\err{ 2}{ 9}
&
             .15\err{ 2}{ 6} &1.55\err{ 1}{ 2} \\
2   &        &  279 & .448\err{ 1}{ 5} & .0152\err{ 1}{ 5} & .175\err{ 3}{11}
&
             .20\err{ 2}{11} &1.97\err{ 1}{ 3} \\
2.67&        &  477 & .426\err{ 2}{ 6} & .0231\err{ 2}{ 5} & .157\err{ 2}{11}
&
             .16\err{ 2}{10} &2.43\err{ 1}{ 3} \\
4   &        & 2111 & .390\err{ 1}{ 4} & .0419\err{ 2}{ 4} & .135\err{ 2}{ 8}
&
             .17\err{ 1}{ 7} &3.28\err{ 1}{ 2} \\
8   &        & 8582 & .319\err{ 6}{11} & .1270\err{11}{18} & .111\err{ 6}{17}
&
             .28\err{ 2}{ 3} &5.70\err{ 2}{ 4} \\
\hline
0   & 2.5115 &  550 & .537\err{ 1}{ 4} & .0337\err{ 2}{ 5} & .233\err{ 2}{ 8}
&
             .26\err{ 1}{ 7} &1.46\err{ 1}{ 1} \\
1   &        & 1320 & .543\err{ 3}{ 7} & .0325\err{ 3}{ 7} & .250\err{ 5}{16}
&
             .20\err{ 3}{10} &1.44\err{ 1}{ 2} \\
1.33&        & 2580 & .513\err{ 1}{ 4} & .0381\err{ 2}{ 4} & .207\err{ 2}{ 7}
&
             .24\err{ 1}{ 8} &1.56\err{ 1}{ 1} \\
2   &        & 1700 & .443\err{ 3}{ 6} & .0643\err{ 3}{ 6} & .142\err{ 6}{13}
&
             .27\err{ 2}{ 6} &2.03\err{ 1}{ 1} \\
\hline
\end{tabular}
\caption{\it Summary of results from fits to the effective potentials
using Eq.~(\protect\ref{fit}) on lattices of size $N_\tau \times 32^3$. The
values of $N_\tau$, which correspond to the temperatures given in the
first column, are described in the text. The third column gives the number
of gauge field configurations used in the analysis.
They are separated by 100 sweeps of overrelaxed
Monte Carlo updates. Remaining autocorrelations have been taken into
account in the error analysis.}
\label{fit_parameter}
\end{center}
\end{table}

\begin{figure}[htb]
\begin{center}
\leavevmode
\epsfxsize=280pt
\epsfbox[20 30 620 600]{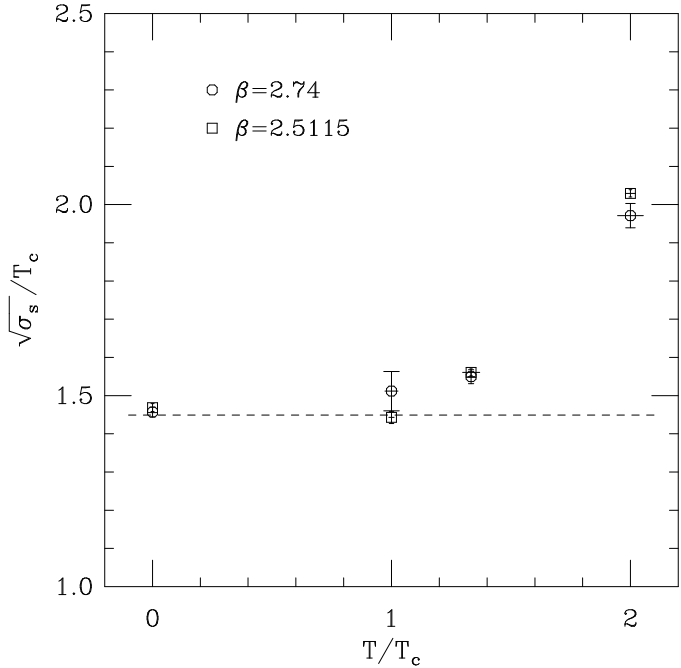}
\end{center}
\caption{\it Square root of the spatial string tension in units of
the critical temperature
versus temperature calculated at two different values of the gauge
coupling. The broken line gives the result for the ratio of the physical
string
tension to the deconfinement temperature averaged over several values
of the critical coupling \protect{\cite{Fingberg}}.
The horizontal error bars indicate the uncertainty in the temperature
scale due to the statistical errors in the determination of the critical
couplings for the deconfinement transition.}
\label{fig:scaling}
\end{figure}

In Fig.~\ref{fig:scaling} we compare the spatial string tension
calculated at $\beta=2.5115$ and $2.74$ at different temperatures.
We find that our data sets are consistent with each other.
Thus, similar to what has been found
for the ratio of the physical string tension to the deconfinement transition
temperature, scaling violations in the ratio
$\sqrt{\sigma_s}/T_c$ are negligible.
This demonstrates that the spatial string tension, indeed, is relevant to
high temperature QCD as it persists in the continuum limit.
Moreover, $\sigma_s$ coincides with the physical,
zero temperature string tension for $T\leq T_c$.

The coupling of the Yang-Mills part of the action of the
effective 3-dimensional theory, $g_3$, derived from a (3+1)-dimensional
$SU(N)$ gauge theory at high temperature, is given in
terms of the temperature and the four-dimensional coupling $g(T)$ as
$g_3^2 = g^2(T) T$. Although the temperature will set the scale also
for other couplings in the 3-dimensional theory, these couplings will in
general have a different dependence on the four-dimensional gauge
coupling $g^2(T)$~\cite{Reisz}. The functional dependence of
$\sigma_s(T)$ on $g^2(T)$ and $T$ thus is not apparent from the general
structure of the effective action. Nonetheless, in a pure
three-dimensional $SU(N)$ gauge theory dimensionful quantities are
proportional to an appropriate power of the three-dimensional coupling
$g_3$. If the temperature dependence of the pure gauge part of the effective
action dominates the spatial string tension we would expect to find
\begin{equation}
\sqrt{\sigma_s (T)} = c g^2(T) T ~~~,
\label{ssttemp}
\end{equation}
where the temperature dependent running coupling constant $g^2(T)$ should,
at high temperatures, be determined by the $\beta$-function of $SU(N)$
in four dimensions.

\begin{figure}[htb]
\begin{center}
\leavevmode
\epsfxsize=280pt
\epsfbox[20 30 620 600]{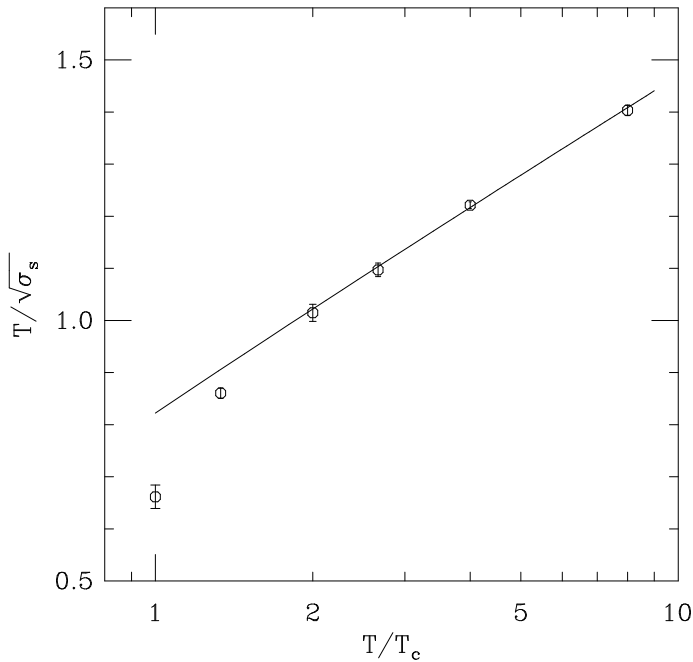}
\end{center}
\caption{\it The ratio of the critical temperature and square root of the
spatial string tension versus temperature for $\beta=2.74$. The line shows
a fit to the data in the region $2\le T/T_c\le 8$ using the 2-loop relation
for $g(T)$ given in Eq.~(\protect{\ref{twoloop}}).}
\label{fig:g3_inv}
\end{figure}

In Fig.~\ref{fig:g3_inv} we have plotted $T/\sqrt{\sigma_s (T)}$
against $T$. From Eq.~(\ref{ssttemp}) this ratio is
expected to be proportional to $g^{-2}(T)$.
We have fitted these data to the two-loop formula for the coupling
in $SU(2)$ gauge theory with the scale parameter $\Lambda_T$,
\begin{equation}
g^{-2}(T) = {11 \over 12 \pi^2} \ln{T/\Lambda_T} +
          {17 \over 44 \pi^2}  \ln(2\ln{T/\Lambda_T})~~~~.
\label{twoloop}
\end{equation}
We find that the temperature dependence of the spatial string tension
is well described by Eqs.~(\ref{ssttemp}) and
(\ref{twoloop}) for temperatures above $2T_c$. From the
two parameter fit to the data shown in Fig.~\ref{fig:g3_inv} in the region
$T \ge 2T_c$ we obtain
\begin{equation}
\sqrt{\sigma_s (T)} = (0.369 \pm 0.014) g^2(T) T ~~~,
\label{sstfit}
\end{equation}
with $\Lambda_T = 0.076(13) T_c$. We note that the
second term in Eq.~(\ref{twoloop}) varies only little with temperature. A
fit with the one loop formula thus works almost equally well; it yields
$\Lambda_T= 0.050(10) T_c$ and $c=0.334(14)$ for the coefficient in
Eq.~(\ref{sstfit}).

It is rather remarkable that the spatial string tension
depends in this simple form on the perturbative $SU(2)$ $\beta$-function
already for $T\ge 2T_c$ and that possible contributions from higher
orders in $g^2(T)$ could be absorbed into the scale parameter $\Lambda_T$.
Moreover, we find that even quantitatively the spatial string tension
agrees well with the string tension of the three-dimensional $SU(2)$ gauge
theory, $\sqrt{\sigma_3} = (0.3340 \pm 0.0025) g_3^2$~\cite{Teper}.
We take this as an indication that, indeed, the spatial string tension is
dominated by the pure gauge part of the effective three-dimensional theory.
We note that the value for $g^2(T)$, determined here from long
distance properties of the (3+1)-dimensional theory, is about a factor two
larger than what has been obtained by comparing the short distance part of
the (3+1)-dimensional heavy quark potential with perturbation
theory~\cite{Lacock}.

\medskip
\noindent
{\bf Acknowledgements:}
The computations have been performed on the Connection Machines at HLRZ, SCRI
and Wuppertal. We thank the staff of these institutes for their support.
FK would like to thank A. Patkos and B. Petersson for discussions on
dimensional reduction at high temperature.
The work of JF and UMH was supported in part by the DOE under grants
\#~DE-FG05-85ER250000 and \#~DE-FG05-92ER40742.
The work of GSB and KS was supported by the EC under grant
\#~SC1*-CT91-0642. We are grateful
to the DFG for supporting the
Wuppertal CM-2 project (grant Schi 257/1-4).

\end{document}